\documentclass{iopconfser}
\usepackage{amsmath}    % for advanced math formatting
\usepackage{amssymb}    % for additional symbols
\usepackage{graphicx}   % for including images (if needed)
\usepackage{bm}         % for bold symbols
\usepackage{physics}    % for physics symbols and formatting
\usepackage{hyperref}
\usepackage{eucal}     % per uno stile diverso di lettere calligrafiche
\usepackage{mathtools} % per strumenti avanzati di formattazione matematica
\usepackage{graphicx}   % per includere immagini
\usepackage{float}      % per il posizionamento delle figure (il comando [H] richiede questo pacchetto)
\usepackage{caption}    % per la gestione delle didascalie (se hai bisogno di personalizzarle ulteriormente)
\usepackage{subcaption} % per avere sottofigure (opzionale, se hai bisogno di gestire più sottofigure in una singola figura)
\usepackage{comment}

\begin{document}

\title{The effect of spacetime torsion on neutrino mixing}

\author{Capolupo A.$^{1}$, Monda S.$^{1}$, Pisacane S.$^{1}$, Quaranta A.$^{2}$ and Serao R.$^{1}$}

\affil{$^1$ Dipartimento di Fisica “E.R. Caianiello” Universit`a di Salerno, and INFN – Gruppo Collegato
	di Salerno, Via Giovanni Paolo II, 132, 84084 Fisciano (SA), Italy}
\affil{$^2$School of Science and Technology, University of Camerino, Via Madonna delle Carceri,
	Camerino, 62032, Italy}

\email{capolupo@sa.infn.it, smonda@unisa.it, gpisacane@unisa.it, aniello.quaranta@unicam.it,
	rserao@unisa.it}

\begin{abstract}
In the framework of quantum field theory, we analyze
the neutrino oscillations in the presence
of a torsion background. We consider the Einstein–
Cartan theory and we study the cases of
constant torsion and of linearly time-dependent
torsion. We derive new neutrino oscillation formulae
which depend on the spin orientation and
the CP-asymmetry formula. Experiment such
as PTOLEMY which analyzes the cosmological
background of neutrino, can provide insights into
the effect shown here \cite{C-Dm-M-Q-S: 2024}.
\end{abstract}

\section{Introduction}
Theories of gravity that extend beyond General Relativity (GR) have a long and complex history \cite{ Capoz1,EX1,EX2,EX3,EX4,EX5,EX6,EX7,EX8,EX9,EX10,EX11,EX12}. These theories have been driven by the need to address the limitations of GR, provide explanations for the dark components of the universe \cite{DM1,DM2,DM3,DM4,DM5,DM6,DM7,DM8,DM9,DE1,DE2,DE3,DE4,DE5,DE6}, and potentially establish a viable framework for the quantisation of gravity. As a result, a large number of such theories have emerged. Some, like the early attempt by Brans and Dicke to incorporate Mach's principle \cite{BransDicke}, involve additional fields beyond the metric \cite{ST1,ST2}. Other theories generalise the Einstein–Hilbert action, often incorporating higher-order curvature invariants \cite{HO1,CU1,CU2}. A natural extension of GR arises when a non-symmetric connection is considered, allowing for the inclusion of torsion \cite{Hehl,Shapiro,CUR1,CUR2,CUR3,CUR4,CUR5,CUR6,CUR7,CUR8,CUR9,CUR10}. Gravitational theories that include torsion may help explain dark matter and dark energy \cite{DETorsion}. Torsion couples naturally to the spin density of matter, resulting in spin-dependent splittings of energy levels \cite{Cabral} and spin oscillations \cite{Cirillo-Lombardo}. Neutrinos, on the other hand, play an important role in both cosmology and astrophysics \cite{NTorsion,ICE1,ICE2,ICE3,ICE4,ICE5,ICE6,ICE7,ICE8,ICE9,ICE10,DUNE1,DUNE2,DUNE3,EXP1,EXP2,EXP3}. Due to their relatively weak interactions and the large quantities in which they are produced, neutrinos serve as a valuable source of information about the universe. They may be related to the original baryon asymmetry \cite{Lepto1}, dark matter \cite{Curv0,Curv0-1}, and dark energy \cite{Kaplan}. Furthermore, neutrinos present numerous challenges to the Standard Model of particle physics, and many aspects of neutrino physics—including the fundamental mechanism behind flavour oscillations \cite{N1,N2,N3,N4,N5-0,N5,N51,N52,N6,N7,N8,N9,N10,N11,N12,N13,N14,N15,N16,N17}, the origin of their mass, and their fundamental nature \cite{NBSM1,NBSM2,NM1,NM2,NM3,NM4,NM5,NM6}—remain unclear.
We analyze the propagation of neutrinos on a
torsion background and study its impact on flavor
oscillations. Neutrino oscillations in the presence
of torsion have been studied in the quantum mechanical
framework. We focus on the simplest
generalization of GR including torsion, the Einstein–
Cartan theory. We consider the cases of
constant torsion and of torsion linearly depending
on time, and we assume that spacetime curvature
is absent.
Our result is: the energy splitting induced by the
torsion term leads to spin-dependent neutrino oscillation
formulae. We also discuss the modifications to the CP asymmetry induced by torsion, which arise as a consequence of the Dirac CP-violating phase in the mixing matrix. Additionally, we show that the CP asymmetry depends on the spin orientations in the presence of the torsion background. It should be noted that in this work, we do not consider the Fermi-like four-fermion interactions among neutrinos that would be induced by integrating out the (non-dynamical) torsion in the Einstein–Cartan framework. Instead, we assume the presence of a background torsion term without specifying its origin. However, this background field can be regarded as the mean torsion field generated by the spin density of a generic fermion field. The Fermi-like four-fermion interactions among neutrinos could affect the neutrino cross-section; however, this is outside the scope of the present study. Moreover, as with general four-fermion interactions \cite{KK}, the interaction induced by torsion may potentially lead to neutrino condensation phenomena \cite{Quaranta-Capolupo}. 

\section{Dirac field quantization with torsion}
We consider the effect of back-ground torsion described by the axial vector field $T^\mu(x)$ on Dirac fields in flat space-time.
We replace the covariant derivative by the ordinary derivative since we are specifically interested in the effects of torsion on Dirac fields. Under these assumptions, the Dirac equation becomes:

\begin{align*}
	i \gamma^\mu \partial_\mu \Psi = m \Psi - \frac{3}{2} T_{\rho} \gamma^\rho \gamma^5 \Psi .
\end{align*}
Setting $u_{\vec{k}}^{r}(t)=e^{-iE^{r}t}u_{\vec{k}}^{r}$ and $v_{\vec{k}}^{r}(t)=e^{iE^{r}t}v_{\vec{k}}^{r}$,
the Dirac field is expanded as usual.
Considering a constant axial torsion directed along the third spatial
axis, the solutions of Dirac equation $u_{\vec{k}}^{r}$ and $v_{\vec{k}}^{r}$ are formally the same as in flat space, except for a spin-dependent mass term $\widetilde{m}^{\pm}=m\pm\frac{3}{2}T^{3}$.
The torsion has indeed the effect of lifting the degeneracy in energy
between the two spin orientations $E_{\vec{k}}^{\pm}=\sqrt{\vec{k}^{2}+\widetilde{m}^{\pm^{2}}}$.

In the case of time-dependent torsion the Dirac equation is formally similar to previous. We use the ansatz
\begin{small}
	\begin{equation*}
		\begin{array}{c}
			u_{\vec{k},\lambda}(t,\boldsymbol{x})=e^{i\boldsymbol{k}\cdot\boldsymbol{x}}\left(\begin{array}{c}
				f_{k}(t)\xi_{\lambda}(\hat{k})\\
				g_{k}(t)\lambda\xi_{\lambda}(\hat{k})
			\end{array}\right)
			\\
			v_{\vec{k},\lambda}(t,\boldsymbol{x})=e^{i\boldsymbol{k}\cdot\boldsymbol{x}}\left(\begin{array}{c}
				g_{k}^{*}(t)\xi_{\lambda}(\hat{k})\\
				-f_{k}^{*}(t)\lambda\xi_{\lambda}(\hat{k})
			\end{array}\right)\,,
		\end{array}
	\end{equation*}
\end{small} for positive energy and 
for negative energy. In the specific case of  $\,\breve{T}^{i} = \alpha^i t$, where $i=1,2,3$ and $\breve{T}^0$ is indipendent on time, one has
\begin{small}
	\begin{equation*}
		\left\{ \begin{array}{c}
			f_{\vec{k},\lambda}(t)=\exp\left\{ -i\frac{t^{2}}{2}\eta\lambda\alpha^{i}\hat{k}^{i}\right\} \exp\left\{ -i\omega_{k,\lambda}t\right\} C_{\vec{k},\lambda}\\
			g_{\vec{k},\lambda}(t)=\frac{k+\eta\lambda\breve{T}^{0}}{\left(\omega_{k,\lambda}+m\right)}\exp\left\{ -i\frac{t^{2}}{2}\eta\lambda\alpha^{i}\hat{k}^{i}\right\} \exp\left\{ -i\omega_{k,\lambda}t\right\} C_{\vec{k},\lambda}\,,
		\end{array}\right.\label{eq: soluzioni f e g}
	\end{equation*}
\end{small}
where \begin{small}
	$\omega_{\pmb{k},\lambda} = \sqrt{m^2 + \left(k + \eta \lambda \breve{T}^0 \right)^2}$, $\eta$	
\end{small} is the coupling parameter, $\lambda$ is the helicity and $\vec{k}$ is the momentum and $C_{\vec{k},\lambda}=\frac{\omega_{k,\lambda}+m}{\left(2\pi\right)^{\frac{3}{2}}\sqrt{\left(\omega_{k,\lambda}+m\right)^{2}+
			\left(k+\eta\lambda\breve{T}^{0}\right)^{2}}}\,.$\\
	
\section{Flavor mixing with torsion}

We denote with $\Psi_{m}^{T}\equiv(\nu_{1},\nu_{2},\nu_{3})$ the free Dirac fields in the presence of torsion.  The flavor fields
are obtained by $\nu_{\sigma}^{\alpha}=\emph{I}_{\theta}^{-1}(t)\nu_{i}^{\alpha}(x)\emph{I}_{\theta}(t)\,,$ where $(\sigma,i)=(e,1),(\mu,2),(\tau,3)$. Indeed the rotation to flavor fields can be recast in terms of the mixing generator $\emph{I}_{\theta}$ where $\emph{I}_{\theta}(t)=\emph{I}_{23}(t)\emph{I}_{13}(t)\emph{I}_{12}(t)\;.$    The action of the mixing generator defines the plane wave
expansion of the flavor fields
\begin{small}
	\begin{equation*}
		\nu_{\sigma}(x)=\sum_{r}\int\frac{d^{3}\boldsymbol{k}}{(2\pi)^{\frac{3}{2}}}\left[u_{\vec{k},i}^{r}\alpha_{\vec{k},\nu_{\sigma}}^{r}(t)+v_{-\vec{k},i}^{r}\beta_{-\vec{k},\nu_{\sigma}}^{r\dagger}(t)\right]\exp\{i\vec{k}\cdot\vec{x}\}\,,
	\end{equation*}
\end{small}

where the flavor annihilators are given by $\alpha_{\vec{k},\nu_{\sigma}}^{r}(t)\equiv\emph{I}_{\theta}^{-1}(t)\alpha_{\vec{k},i}^{r}\emph{I}_{\theta}(t)\,,
\quad\beta_{-\vec{k},\nu_{\sigma}}^{r\dagger}(t)\equiv\emph{I}_{\theta}^{-1}(t)\beta_{-\vec{k},i}^{r\dagger}(t)\emph{I}_{\theta}(t)\,.$
They annihilate the flavor vacuum $\alpha_{\vec{k},\nu_{\sigma}}^{r}\left|0\right\rangle _{f}=0=\beta_{-\vec{k},\nu_{\sigma}}^{r}\left|0\right\rangle _{f}$.
The explicit relations of the the flavor annihilators, for $\vec{k}=(0,0,\left|\vec{k}\right|)$, are\vspace{-10pt} 

	\begin{align*}
		\alpha_{\vec{k},\nu_{e}}^{r}(t) & =c_{12}c_{13}\alpha_{\vec{k},1}^{r}+s_{12}c_{13}\left(\left(\Gamma_{12;\vec{k}}^{rr}(t)\right)^{*}\alpha_{\vec{k},2}^{r}+
		\varepsilon^{r}\left(\Sigma_{12;\vec{k}}^{rr}(t)\right)\beta_{-\vec{k},2}^{r\dagger}\right)\,,
	\end{align*}
and similar relations for other annihilators. 
Bogoliubov coefficients $ \Gamma_{ij;\vec{k}}^{rr}$ and $ \Sigma_{ij;\vec{k}}^{rr}$, appearing in the expressions of  the
flavor annihilators, are given by the inner product of the solutions of Dirac equations with different masses. In order to distinguish
the case of constant torsion from that of time-dependent torsion, we use notation
$ \Gamma_{ij;\vec{k}}^{rr} = \Xi_{ij;\vec{k}}^{rr} $ and
$ \Sigma_{ij;\vec{k}}^{rr} = \chi_{ij;\vec{k}}^{rr} $
for constant torsion and $ \Gamma_{ij;\vec{k}}^{rr} = \Pi_{ij;\vec{k}}^{rr} $ and $ \Sigma_{ij;\vec{k}}^{rr} = \Upsilon_{ij;\vec{k}}^{rr} $ for time-dependent torsion. \subsection{Bogoliubov Coefficients with Constant Torsion}

For constant torsion, the modules of the  Bogoliubov coefficients are given by
\vspace{-6pt}\[
\left|\Xi_{i,j;\vec{k}}^{r,s}\right|\equiv u_{\vec{k},i}^{r\text{\ensuremath{\dagger}}}u_{\vec{k},j}^{s}=v_{-\vec{k},i}^{s\text{\ensuremath{\dagger}}}v_{-\vec{k},j}^{r}\,,
\qquad\left|\chi_{i,j;\vec{k}}^{r,s}\right|\equiv\varepsilon^{r}u_{\vec{k},1}^{r\text{\ensuremath{\dagger}}}
v_{-\vec{k},2}^{s}=-\varepsilon^{r}u_{\vec{k},2}^{r\text{\ensuremath{\dagger}}}v_{-\vec{k},1}^{s}\,.
\]

Notice that, in reference frame $\vec{k}=(0,0,\left|\vec{k}\right|)$, $\Xi_{i,j;\vec{k}}^{r,s}$ and $\chi_{i,j;\vec{k}}^{r,s}$ vanish for $r\neq s$. Explicitly, we have
\vspace{-6pt}\[
\begin{array}{c}
	\Xi_{ij;\vec{k}}^{\pm\pm}=N_{i}^{\pm}N_{j}^{\pm}\left[1+\frac{k^{2}}{\left(E_{\vec{k},i}^{\pm}+\widetilde{m}_{i}^{\pm}\right)
		\left(E_{\vec{k},j}^{\pm}+\widetilde{m}_{j}^{\pm}\right)}\right]=\cos(\xi_{ij;\vec{k}}^{\pm \pm})\,,\\
	\chi_{ij;\vec{k}}^{\pm\pm}=N_{i}^{\pm}N_{j}^{+}\left[\frac{k_{3}}{E_{\vec{k},j}^{\pm}+\widetilde{m}_{j}^{\pm}}-\frac{k_{3}}{E_{\vec{k},i}^{\pm}+
		\widetilde{m}_{i}^{\pm}}\right]=\sin(\xi_{ij;\vec{k}}^{\pm\pm})\,,
\end{array}
\]
with the spin-dependent masses and the normalization coefficients
given explicitly by $\widetilde{m}_{i}^{\pm}\equiv m_{i}\pm\frac{3}{2}T^{3}$
and $N_{i}^{\pm}=\frac{\sqrt{E_{\vec{k},i}^{\pm}+\widetilde{m}_{i}^{\pm}}}{\sqrt{2E_{\vec{k},i}^{\pm}}}$,
respectively. The sign factor is defined as $\varepsilon^{\pm}=\mp1$.
Additionally%MDPI: Please carefully recheck the accuracy of this symbol \mp.
, $(E_{\vec{k},i}^{\pm})^{2}=\vec{k}^{2}+(\widetilde{m}_{i}^{\pm})^{2}$
and $\xi_{ij;\vec{k}}^{\pm\pm}=\arctan\left(\frac{\left|V_{ij;\vec{k}}^{\pm\pm}\right|}{\left|U_{ij;\vec{k}}^{\pm\pm}\right|}\right)$.
The canonicity of the Bogoliubov transformations
is ensured by relations
$
\sum_{r}\left(\left|\Xi_{ij;\vec{k}}^{\pm r}\right|^{2}+\left|\chi_{ij;\vec{k}}^{\pm r}\right|^{2}\right)=1
$
where $i,j=1,2,3$ and $j>i$. Moreover, the time dependence of $\Xi_{ij;\vec{k}}^{\pm r}$ and $\chi_{ij;\vec{k}}^{\pm r}$  is expressed by
\vspace{-6pt}\[
\Xi_{ij;\vec{k}}^{rs}(t)=\left|\Xi_{ij;\vec{k}}^{rs}\right|e^{i\left(E_{\vec{k},j}^{s}-
	E_{\vec{k},i}^{r}\right)t}\;,\:\;\;\;\;\chi_{ij;\vec{k}}^{rs}(t)=\left|\chi_{ij;\vec{k}}^{rs}\right|e^{i\left(E_{\vec{k},j}^{s}+
	E_{\vec{k},i}^{r}\right)t}\,.
\]

\subsection{Bogoliubov Coefficients with Time-Dependent Torsion}

%Similarly to the previous case, a mixing generator is introduced that
%binds the flavour fields to the free ones. \foreignlanguage{italian}{The
	%relations of flavour annihilators in terms of those with a defined
	%mass are analogous to those given in the appendix with the Bogoliubov
	%coefficients replaced by those in Eqs.$(\ref{eq: Bogoljubov U++ time-dependent})$,$(\ref{eq: Bogoljubov V++ time-dependent})$.}
In this case, the Bogoliubov coefficients are denoted with $\varPi_{ij;\vec{k}}^{rs}(t)=\left(u_{\vec{k},i}^{r},u_{\vec{k},j}^{s}\right)_{t}$
and $\varUpsilon_{ij;\vec{k}}^{rs}(t)=\left(u_{\vec{k},i}^{r},v_{\vec{k},j}^{s}\right)_{t}$.
The mixed coefficients are zero, and explicitly, we have\vspace{-6pt} 
\begin{align*}
	\varPi_{ij;\vec{p}}^{ss}(t) & = \left(2\pi\right)^{3}\exp\left\{ -i\left(\omega_{p,s}^{j}-\omega_{p,s}^{i}\right)t\right\} \left(C_{\vec{p},i}^{s}\right)^{*}\left(C_{\vec{p},j}^{s}\right)\left[1 + \frac{\left|p + s\eta\breve{T}^{0}\right|^{2}}{\left(\omega_{p,s}^{i}+m_{i}\right)\left(\omega_{p,s}^{j}+m_{j}\right)}\right] \,, \\
	\varUpsilon_{ij;\vec{p}}^{ss}(t) & = \left(2\pi\right)^{3}\exp\left\{ +it^{2}\eta\alpha^{i}\hat{p}^{i}\right\} \exp\left\{ +i\left(\omega_{p,s}^{j} + \omega_{p,s}^{i}\right)t\right\} \left(C_{\vec{p},i}^{s}\right)^{*}\left(C_{\vec{p},j}^{s}\right)^{*}\left(p + s\eta\breve{T}^{0}\right) \left[\frac{1}{\omega_{p,+}^{j}+m_{j}} - \frac{1}{\omega_{p,+}^{i}+m_{i}}\right] \,,
\end{align*}

where $s=\pm$, $i,j=1,2,3$ and $j>i$. 
	In the ultrarelativistic case ($p \gg m_j$), one has:
	\vspace{-6pt}\[
	\varPi_{\vec{p}}^{rr}(t)\longrightarrow1\,,\;\hfill\varUpsilon_{\vec{p}}^{rr}(t)\longrightarrow0\,
	\]
	for any $t$. Moreover, in the absence of torsion (i.e., $\breve{T}^{\mu}=0$)
	these coefficients coincide  with those presented in the Minkowski metric.
The canonicity of the Bogoliubov transformations
is satisfied by the following relations:
$
\sum_{r}\left(\left|\varPi_{ij;\vec{k}}^{\pm r}\right|^{2}+\left|\varUpsilon_{ij;\vec{k}}^{\pm r}\right|^{2}\right)=1\,.
$

\section{ Neutrino oscillations with background torsion}
By analyzing flavor currents and charges we define the flavor charger in presence of torsion as \\
$::\,Q_{\nu_{\sigma}}\,::=\sum_{r}\int d^{3}\boldsymbol{k}\left(\alpha_{\vec{k},\nu_{\sigma}}^{r\dagger}(t)\alpha_{\vec{k},\nu_{\sigma}}^{r}(t)-
\beta_{\vec{k},\nu_{\sigma}}^{r\dagger}(t)\beta_{\vec{k},\nu_{\sigma}}^{r}(t)\right),$ with $\sigma=e,\mu,\tau
$
and, $\left.::\cdots::\right.$, denoting the normal ordering with respect
to the flavor vacuum state $\left|0\right\rangle _{f}$. The oscillation formulas are spin depending because of the spin dependence of Bogoliubov coefficients (ruling the amplitudes) and the frequences. %On the other hand 

\begin{figure}[H]
	\begin{centering}
		\includegraphics[scale=0.6]{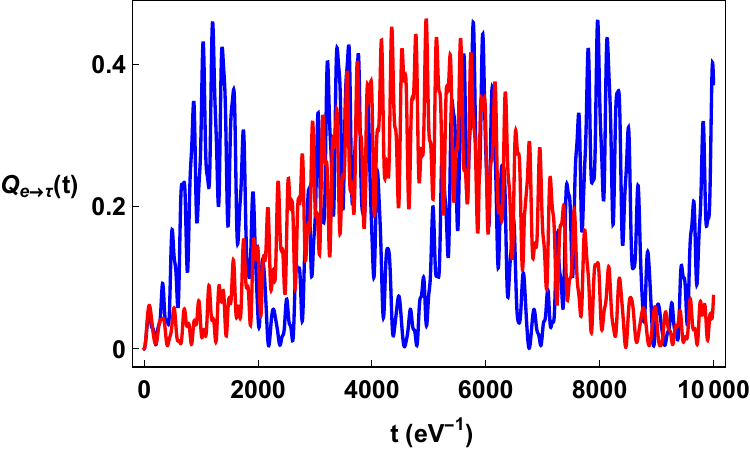}\includegraphics[scale=0.6]{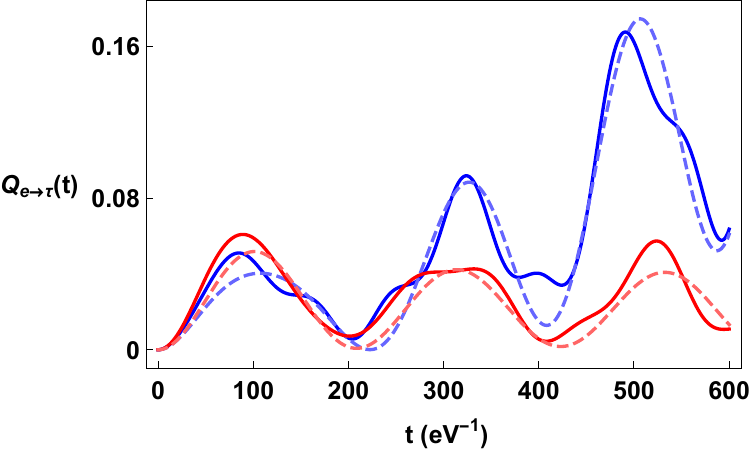}
		\par\end{centering}
	\caption{\label{fig:Pe-mu. 3 generazioni. z50}In the left-hand panel, plot of   $\mathcal{Q^{\uparrow}}_{\nu_{e}\rightarrow\nu_{\tau}}^{\vec{k}}(t)$
		(blue line) and $\mathcal{Q^{\downarrow}}_{\nu_{e}\rightarrow\nu_{\tau}}^{\vec{k}}(t)$
		(red line) as a function of time.  In the right
		panel, detail of the same formulae and comparison with the corresponding QM
		formulae (dashed line).}
\end{figure}

\begin{figure}[htbp]
	\centering
	\includegraphics[scale=0.6]{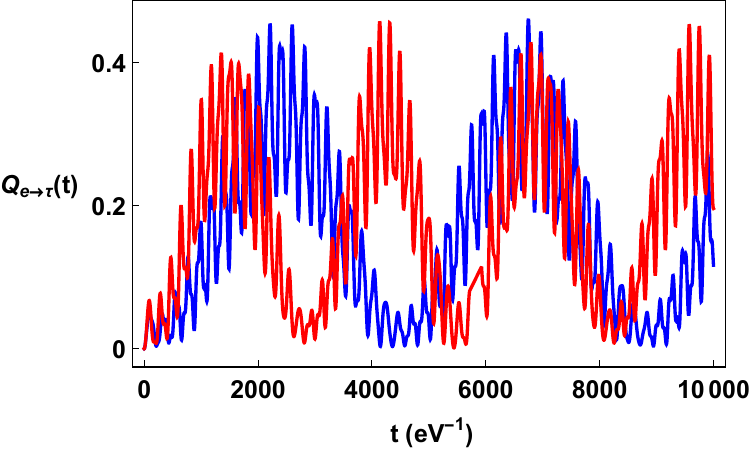} 
	\includegraphics[scale=0.6]{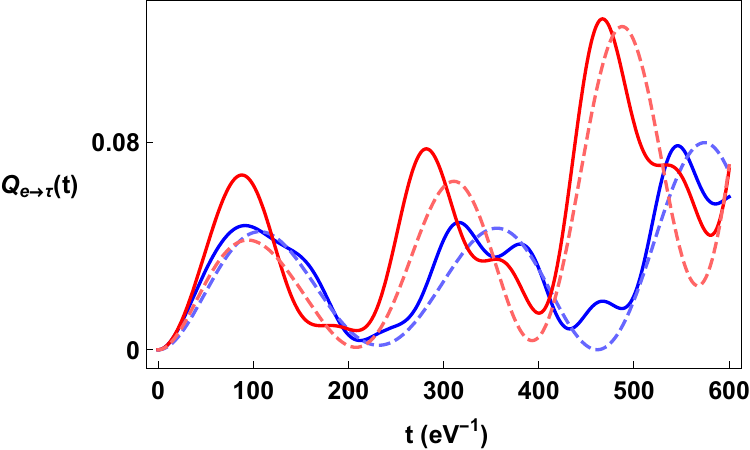}
	\caption{\label{fig:petautime3sapori}In the left-hand panel, plot of $\mathcal{Q^{\uparrow}}_{\nu_{e}\rightarrow\nu_{\tau}}^{\vec{k}}(t)$
		(blue line) and $\mathcal{Q^{\downarrow}}_{\nu_{e}\rightarrow\nu_{\tau}}^{\vec{k}}(t)$
		(red line) as a function of time. In the right panel, the detail of the same formulae and comparison with the corresponding
		QM formulae (dashed line).}
\end{figure}

%where the Bogoliubov coefficients  and the frequencies are spin depending.
%Then, the oscillation formulae are highly spin-dependent,  %$\mathcal{Q^{\uparrow}}_{\nu_{\sigma}\rightarrow\nu_{\rho}}^{\vec{k}}(t)\neq
%\mathcal{Q^{\downarrow}}_{\nu_{\sigma}\rightarrow\nu_{\rho}}^{\vec{k}}(t)$.
On the other hand in the QM formulae the spin orientation affects only the frequencies. We report the transition formulas for sample values of costant torsion and time dependent torsion. 
We report the transition formulae for sample values of torsion and momentum. We consider values of neutrino
masses $m_{1}\approx10^{-3}~\mathrm{eV}$, $m_{2}\approx9\times10^{-3}~\mathrm{eV}$,
and $m_{3}\approx2\times10^{-2}~\mathrm{eV}$, in order that $\Delta m_{12}^{2}\approx7.56\times10^{-5}~\mathrm{eV}^{2}$
and $\Delta m_{23}^{2}\approx2.5\times10^{-3}~\mathrm{eV}^{2}$, and
of mixing angles such that $\sin^{2}(2\theta_{13})=0.10,$ $\sin^{2}(2\theta_{23})=0.97,$
and $\sin^{2}(2\theta_{12})=0.861$, which are compatible with the experimental data. We also consider $\delta=\pi/4$,
and a  fixed value of momentum $k\simeq2\times10^{-2}~\mathrm{eV}$
and of torsion $|T^{3}|\simeq2\times10^{-4}~\mathrm{eV}$.
In the following figure are plotted $\mathcal{Q^{\uparrow}}_{\nu_{e}\rightarrow\nu_{\tau}}^{\vec{k}}(t)$ and $\mathcal{Q^{\downarrow}}_{\nu_{e}\rightarrow\nu_{\tau}}^{\vec{k}}(t)$ as function of time in the specific case of costant torsion in Fig \ref{fig:Pe-mu. 3 generazioni. z50} and time dependent torsion in Fig \ref{fig:petautime3sapori}; they are compared with the corrisponding QM formulae (right panel). \\

As in the general case, quantum field theoretical effects on neutrino oscillations are
relevant at low momenta and this remains true also in the presence of torsion. The quantum field theoretical effects may affect the
capture rate of experiments dealing with low-energy neutrinos, such as Ptolemy. Other
experiments, such as DUNE [107,108], which feature much more energetic neutrinos, are
essentially unaffected by quantum field theoretical effects.

\section{\textbf{\emph{CP} }Violation and Flavor Vacuum}\label{sec:5}

We now  study the impact of torsion on the $CP$ violation in neutrino oscillation due to the presence of Dirac phase
in the mixing matrix. For fixed spin orientation, say $\uparrow$,  the $CP$ asymmetry $\Delta_{\uparrow;CP}^{\rho\sigma}$ can be defined in QFT,
in a similar way to what was achieved in \linebreak ref. \cite{N5}; and then,
$
\Delta_{\uparrow;CP}^{\rho\sigma}(t)  \equiv\mathcal{Q^{\uparrow}}_{\nu_{\rho}\rightarrow\nu_{\sigma}}^{\vec{k}}(t)+
\mathcal{Q^{\uparrow}}_{\overline{\nu}_{\rho}\rightarrow\overline{\nu}_{\sigma}}^{\vec{k}}(t)\,,
$
where $\rho,\sigma=e,\mu,\tau$.  Notice that
a $+$ sign appears in front of the probabilities for the antineutrinos
in place of   $-$, because the antineutrino states already
carry a negative flavor charge $Q_{\sigma}$.
%
%Recalling that the following
%relations are satisfied: $\sum_{\sigma}Q_{\nu_{\sigma}}(t)=Q\,,$
%$\left\langle \nu_{\rho}\right|Q\left|\nu_{\rho}\right\rangle =1\,,$
%and $\left\langle \overline{\nu}_{\rho}\right|Q\left|\overline{\nu}_{\rho}\right\rangle =-1\,,$
%it is easy to show that $\sum_{\sigma}\Delta_{\uparrow;CP}^{\rho\sigma}=0\,,$
%for $\rho,\sigma=e,\mu,\tau\,.$
For $\nu_{e}\rightarrow\nu_{\mu}$ transition, with $r~=~\uparrow,\downarrow$, the $CP$ asymmetry is explicitly

	%\centering %%\fulllength If there is a figure in wide page, please release command \centering
	\begin{align}
		\Delta_{r;CP}^{e\mu}(t) & =4J_{CP}\left[\left|\Gamma_{12;\vec{k}}^{\pm \pm}\right|^{2}\sin\left(2\Delta_{12;\vec{k}}^{\pm}t\right)-
		\left|\Sigma_{12;\vec{k}}^{\pm \pm}\right|^{2}\sin\left(2\Omega_{12;\vec{k}}^{\pm}t\right)\right.
		+\left(\left|\Gamma_{12;\vec{k}}^{\pm \pm}\right|^{2}-
		\left|\Sigma_{13;\vec{k}}^{\pm \pm}\right|^{2}\right)\sin\left(2\Delta_{23;\vec{k}}^{\pm}t\right)\nonumber \\
		&+
		\left(\left|\Sigma_{12;\vec{k}}^{\pm \pm}\right|^{2}-\left|\Sigma_{13;\vec{k}}^{\pm \pm}\right|^{2}\right)\sin\left(2\Omega_{23;\vec{k}}^{\pm}t\right)
		\left.-\left|\Gamma_{13;\vec{k}}^{\pm \pm}\right|^{2}\sin\left(2\Delta_{13;\vec{k}}^{\pm}t\right)+
		\left|\Sigma_{13;\vec{k}}^{\pm \pm}\right|^{2}\sin\left(2\Omega_{13;\vec{k}}^{\pm}t\right)\right]\,,\label{eq: CPup}
	\end{align}

where one has to consider  $\Gamma_{ij;\vec{k}}^{++}$ and $ \Sigma_{ij;\vec{k}}^{++}$ for spin up and
$\Gamma_{ij;\vec{k}}^{--}$ and $ \Sigma_{ij;\vec{k}}^{--}$ for spin down.
One also has $\Delta_{r;CP}^{e\tau}(t)=-\Delta_{r;CP}^{e\mu}(t)$ with
$r=~\uparrow,\downarrow$.
Remarkably, the presence of torsion induces a  $CP$ asymmetry depending on spin orientation.

Furthermore, we  make some observations on the condensate structure of the flavor vacuum in the presence of torsion. In this case,
$\ket{0_{f}(t)}$   breaks the spin symmetry, resulting in a different
condensation density for particles of spin up and down.
Such  densities are evaluated by computing the expectation
values of the number operators for free fields
$N_{\alpha_{j},\vec{k}}^{r} = \alpha_{\vec{k},j}^{r\dagger}\alpha_{\vec{k},j}^{r}$
and $N_{\beta_{j},\vec{k}}^{r}=\beta_{\vec{k},j}^{r\dagger}\beta_{\vec{k},j}^{r}$,
on $\ket{0_{f}(t)}$.
One has\vspace{-6pt} 

	%\centering %%\fulllength If there is a figure in wide page, please release command \centering
	\begin{align}
		\mathcal{N}_{1;\vec{k}}^{r} & =_{f}\left\langle 0(t)\right|\text{\ensuremath{N_{\alpha_{1},\vec{k}}^{r}}}\left|0(t)\right\rangle _{f}={}_{f}\left\langle 0(t)\right|\text{\ensuremath{N_{\beta_{1},\vec{k}}^{r}}}\left|0(t)\right\rangle _{f}  = s_{12}^{2}c_{13}^{2}\left|\Sigma_{12;\vec{k}}^{\pm \pm}\right|^{2}+s_{13}^{2}\left|\Sigma_{13;\vec{k}}^{\pm \pm}\right|^{2}\,,\label{eq:condensato m1}
		\\
		\mathcal{N}_{2;\vec{k}}^{r} & =_{f}\left\langle 0(t)\right|\text{\ensuremath{N_{\alpha_{2},\vec{k}}^{r}}}\left|0(t)\right\rangle _{f}={}_{f}\left\langle 0(t)\right|\text{\ensuremath{N_{\beta_{2},\vec{k}}^{r}}}\left|0(t)\right\rangle _{f}
		=\left|-s_{12}c_{23}+e^{i\delta}c_{12}s_{23}s_{13}\right|^{2}\left|\Sigma_{12;\vec{k}}^{\pm \pm}\right|^{2}+s_{23}^{2}c_{13}^{2}
		\left|\Sigma_{23;\vec{k}}^{\pm \pm}\right|^{2}\,,\label{eq:condensato m2}
		\\ \nonumber
		\mathcal{N}_{3;\vec{k}}^{r} & =_{f}\left\langle 0(t)\right|\text{\ensuremath{N_{\alpha_{3},\vec{k}}^{r}}}\left|0(t)\right\rangle _{f}={}_{f}\left\langle 0(t)\right|\text{\ensuremath{N_{\beta_{3},\vec{k}}^{r}}}\left|0(t)\right\rangle _{f} \\
		&=\left|-c_{12}s_{23}+e^{i\delta}s_{12}c_{23}s_{13}\right|^{2}\left|\Sigma_{23;\vec{k}}^{\pm \pm}\right|^{2}+
		\left|s_{12}s_{23}+e^{i\delta}c_{12}c_{23}s_{13}\right|^{2}\left|\Sigma_{13;\vec{k}}^{\pm \pm}\right|^{2}\,\ ,\label{eq: condensato m3}
	\end{align}

where,
$r=~\uparrow,\downarrow$.

It is important to note that flavor vacuum condensation is a general consequence of the quantum field theory treatment of neutrino mixing \cite{N5-0, N5}. In the presence of torsion, the novelty is represented by the spin orientation dependent on such condensation yielding, as shown in the above equation, distinct condensation density for different spin orientations.  The physical implications of the flavor condensation are represented by correction to the amplitude of the neutrino oscillation formulae and by possible contributions to the dark components of the Universe \cite{Curv0, Curv0-1, Kaplan}.

\section{Conclusions}

In this work, we have examined the Einstein--Cartan theory and studied neutrino propagation in the presence of torsion within the framework of quantum field theory. Through this analysis, we derived new oscillation formulas that are dependent on the spin orientations of the neutrino fields. Specifically, we demonstrated that the energy splitting induced by the torsion term influences both the oscillation frequencies and the Bogoliubov coefficients, which represent the amplitudes of the oscillation formulas. We considered flat space-time and explored two distinct types of torsion terms: constant torsion and linearly time-dependent torsion.

Both cases exhibit the following general behaviour: the spin dependence of oscillations is most pronounced for values of torsion comparable to the neutrino momentum and mass. For significantly larger torsion values, the flavour oscillations become almost independent of spin. Furthermore, sufficiently large torsion can effectively suppress flavour oscillations. These behaviours are also observed in the $CP$ asymmetry.

The effects of torsion on neutrino oscillations are particularly significant in non-relativistic regimes. Therefore, future experiments involving neutrinos with very low momenta, such as PTOLEMY, could provide an opportunity to verify these results.

% If authors have biography, please use the format below
%\section*{Short Biography of Authors}
%\bio
%{\raisebox{-0.35cm}{\includegraphics[width=3.5cm,height=5.3cm,clip,keepaspectratio]{Definitions/author1.pdf}}}
%{\textbf{Firstname Lastname} Biography of first author}
%
%\bio
%{\raisebox{-0.35cm}{\includegraphics[width=3.5cm,height=5.3cm,clip,keepaspectratio]{Definitions/author2.jpg}}}
%{\textbf{Firstname Lastname} Biography of second author}

% For the MDPI journals use author-date citation, please follow the formatting guidelines on http://www.mdpi.com/authors/references
% To cite two works by the same author: \citeauthor{ref-journal-1a} (\citeyear{ref-journal-1a}, \citeyear{ref-journal-1b}). This produces: Whittaker (1967, 1975)
% To cite two works by the same author with specific pages: \citeauthor{ref-journal-3a} (\citeyear{ref-journal-3a}, p. 328; \citeyear{ref-journal-3b}, p.475). This produces: Wong (1999, p. 328; 2000, p. 475)

%%%%%%%%%%%%%%%%%%%%%%%%%%%%%%%%%%%%%%%%%%
%% for journal Sci
%\reviewreports{\\
	%Reviewer 1 comments and authors’ response\\
	%Reviewer 2 comments and authors’ response\\
	%Reviewer 3 comments and authors’ response
	%}
%%%%%%%%%%%%%%%%%%%%%%%%%%%%%%%%%%%%%%%%%%

\end{document}